\documentclass[aps,prb,twocolumn,showpacs]{revtex4}
\usepackage{epsfig}
\usepackage{times}
\usepackage{amsmath}
\begin{document}

\title{Delay-induced multiple stochastic resonances on scale-free neuronal networks}

\author{Qingyun Wang,$^{\star,\dagger}$ Matja{\v z} Perc,$^\ddagger$ Zhisheng Duan,$^\star$ and Guanrong Chen$^{\star,\S}$}

\affiliation
{$^\star$State Key Laboratory for Turbulence and Complex Systems, Department of \\ Mechanics and Aerospace Engineering, College of Engineering, Peking University, Beijing 100871, China \\
$^\dagger$School of Statistics and Mathematics, Inner Mongolia Finance and Economics College, Huhhot 010051, China \\
$^\ddagger$Department of Physics, Faculty of Natural Sciences and Mathematics, University of \\ Maribor, Koro{\v s}ka cesta 160, SI-2000 Maribor, Slovenia \\
$^\S$Department of Electronic Engineering, City University of Hong Kong, Hong Kong SAR, China}

\begin{abstract}
We study the effects of periodic subthreshold pacemaker activity and time-delayed coupling on stochastic resonance over scale-free neuronal networks. As the two extreme options, we introduce the pacemaker respectively to the neuron with the highest degree and to one of the neurons with the lowest degree within the network, but we also consider the case when all neurons are exposed to the periodic forcing. In the absence of delay, we show that an intermediate intensity of noise is able to optimally assist the pacemaker in imposing its rhythm on the whole ensemble, irrespective to its placing, thus providing evidences for stochastic resonance on the scale-free neuronal networks. Interestingly thereby, if the forcing in form of a periodic pulse train is introduced to all neurons forming the network, the stochastic resonance decreases as compared to the case when only a single neuron is paced. Moreover, we show that finite delays in coupling can significantly affect the stochastic resonance on scale-free neuronal networks. In particular, appropriately tuned delays can induce multiple stochastic resonances independently of the placing of the pacemaker, but they can also altogether destroy stochastic resonance. Delay-induced multiple stochastic resonances manifest as well-expressed maxima of the correlation measure, appearing at every multiple of the pacemaker period. We argue that fine-tuned delays and locally active pacemakers are vital for assuring optimal conditions for stochastic resonance on complex neuronal networks.
\end{abstract}

\maketitle

\textbf{It is well known that noise can play a constructive role in different types of nonlinear dynamical systems, and stochastic resonance is perhaps the most prominent example of this fact. The objective of this article is to extend the scope of stochastic resonance to complex networks, whereby the deterministic periodic input is not only limited in its strength but also its outreach. More precisely, scale-free neuronal networks are studied on which the subthreshold periodic forcing is introduced only to a single neuron of the network, thus acting as a pacemaker. We want to determine to what extent the complex scale-free topology can aid the pacemaker to entrain the complete neuronal ensemble with the help of fine-tuned additive noise. Moreover, the new findings are compared with results obtained via the more traditional setup where every neuron of the network is subjected to a weak periodic forcing. It is found that scale-free topologies are very efficient in propagating noise-supported localized weak rhythmic activities. Also, it is found that these paced networks are superior to globally forced networks in that stochastic resonance is better expressed on the former. Importantly, since time delays are inherent to the nervous system we take this explicitly into account via time-delayed coupling. We report on the occurrence of delay-induced multiple stochastic resonances on scale-free neuronal networks, which appear due to the locking between the delay length and the oscillation period of the pacemaker.}

\section{Introduction}

The constructive role of noise in nonlinear dynamics has been the subject of intensive research in the past, and it remains a vibrant topic today. Especially the phenomenon of stochastic resonance \cite{do1,do2,do4} has been studied extensively due to its applications in many different fields, ranging from physical to social systems. In general, stochastic resonance is characterized by the optimization of the output signal-to-noise ratio in a nonlinear dynamical system following the addition of a weak external signal. \cite{sr1,sr2,sr3,sr4,sr5} Coherence resonance is a closely related phenomenon, which however, refers to the resonant response of a dynamical system to pure noise, hence it has also been termed as autonomous stochastic resonance. \cite{cr1,cr2,cr3}. Notably, both stochastic and coherence resonances have been reported existing in a wide variety of neuronal models. \cite{sr1,sr2,cr2,cr3} Of particular interest for the present work, due to the usage of the same mathematical model of neuronal dynamics, multiple resonances in the Rulkov neuronal model \cite{r20} have also been observed and reported. \cite{mtcr3} Effects of correlated noise on the dynamics of coupled neurons have also been investigated thoroughly, \cite{xfx1,xfx2,xfx3,xfx4,xfx5} whereby the importance of different temporal and spatial correlation lengths has been firmly established.

Recently, stochastic and coherence resonances on complex neuronal networks have attracted increasing attention, and indeed some new features due to the underlying complex interaction topologies have been revealed. The stochastic resonance on excitable Watts-Strogatz small-world networks \cite{ws} via a pacemaker was studied by means of the discrete Rulkov map, \cite{mp1} where it was shown that only for intermediate coupling strengths is the small-world property able to enhance the stochastic resonance. Moreover, the stochastic resonance on Newman-Watts networks \cite{nw} of Hodgkin-Huxley neurons could be amplified via fine-tuning of the small-world network structure, depending significantly on the coupling strengths among neurons and the driving frequency of the pacemaker. \cite{mp2} Coherence resonance on Watts-Strogatz small-world Hodgkin-Huxley neuronal networks has also been investigated, and it was found that increasing the randomness of the network topology leads to enhancement of temporal coherence. \cite{com1} Furthermore, Kwon \textit{et al.} \cite{com2} showed that the coherence resonance can be considerably improved by adding a small fraction of long-range connections for an intermediate coupling strength in a Watts-Strogatz small-world neuronal network with spatially correlated noise input. Preceding these studies were reports on the array-enhanced coherence resonance by using an array of coupled FitzHugh-Nagumo neurons, \cite{xfx1,arr1} as well as stochastic resonance on small-world networks of overdamped bistable oscillators. \cite{swsr1} Noise-induced phenomena in two-dimensional spatially extended neuronal networks have also attracted considerable attention in the past, and several studies were devoted to the exploration of possible effects of noise. There are some comprehensive reviews dedicated to this subject. \cite{SA07,GO99} In particular, while spatial coherence resonance was first introduced near pattern forming instabilities, \cite{add1} it subsequently was reported also in excitable neuronal media. \cite{MP1} Noteworthy, spatiotemporal coherence of noise-induced patterns has also been investigated on a regular Hodgkin-Huxley neuronal network, \cite{or22} and it was found that the order of the firing rate function could be enhanced as the connections amongst neurons became stronger.

In this study, the objective is to extend the scope of stochastic resonance on complex neuronal networks, particularly scale-free networks, \cite{r21} pacemakers, and time-delayed coupling, thus bringing the setup closer to actual conditions. Delays are inherent to the nervous system because of the finite speed at which action potentials propagate across neuron axons, and due to time lapses occurring in both dendritic and synaptic processing. \cite{or23} Notably, it has been suggested that time delays can facilitate neural synchronization and lead to many interesting and even unexpected dynamical phenomena. \cite{or24,or25,or26} Moreover, since a power-law distribution of the degrees of neurons has been found applicable for the coherence among activated voxels using functional magnetic resonance imaging, \cite{cn2} and the robustness against simulated lesions of anatomic cortical networks was also found to be very similar to that of a scale-free network, \cite{sm2} our study addresses a relevant system setup. We report on the pacemaker-driven stochastic resonance in scale-free neuronal networks, and compare it with the more classical setup entailing the subthreshold periodic forcing of all neurons constituting the network. Interestingly, we find that stochastic resonance with the locally acting pacemaker is better expressed than the one with the globally forced neuronal network. Primarily though, we present some non-trivial effects induced by finite delays in coupling. In particular, we show that multiple stochastic resonances can occur on scale-free neuronal networks if the duration of the delay is appropriately tuned. This is primarily attributed to the emergence of locking between the delay length and the oscillation period of the pacemaker.

The remainder of this paper is organized as follows. In the next section the Rulkov map model is reviewed, \cite{r20} which will be employed to obtain an efficient setup for simulating neuronal dynamics on scale-free networks. \cite{r21} In Section II, the time-delayed coupling scheme and the measure for stochastic resonance are introduced as well. Main results are presented in Section III, whereas the last Section summarizes the new findings and discusses their implications.

\section{Mathematical model and setup}

To effectively simulate the neuronal dynamics on scale-free networks, the Rulkov map is utilized, \cite{r20} which succinctly captures some of the main dynamical features of the complex continuous-time models yet ensures exceptionally efficient numerical analysis and processing. The spatiotemporal evolution of the studied neuronal network is governed by the following iteration equations:
\begin{eqnarray}
x^{(i)}(n+1)&=& \alpha f[x^{(i)}(n)]+y^{(i)}(n)+\sigma
\xi^{(i)}(n)\nonumber\\*
&&+ D\sum_{j}\varepsilon^{{i,j}}\left[x^{j}(n-\tau)-x^{i}(n)\right], \\
y^{(i)}(n+1)&=& y^{(i)}(n)-\beta x^{(i)}(n)-\gamma, \ i=1,\ldots,N
\nonumber
\end{eqnarray}
Here, $x^{(i)}(n)$ is the membrane potential of the $i$-th neuron and $y^{(i)}(n)$ is the variation of ion concentration, representing the fast and the slow variables, respectively. The slow temporal evolution of $y^{(i)}(n)$ is due to the small values of the positive parameters $\beta$ and $\gamma$, which within this study are chosen as $\beta = \gamma = 0.001$ unless otherwise stated. Moreover, $n$ is the discrete-time index, while $\alpha$ is the main parameter determining the dynamics of individual neurons constituting the scale-free network. If $\alpha < 2.0$ then all neurons occupy excitable fixed (equilibrium) points $[x^* = -1,y^* = -1-(\alpha/2)]$, whereas if $\alpha > 2.0$ then regular bursting oscillations (limit cycles in the phase space) emerge via a Hopf bifurcation. \cite{r20} For larger $\alpha$ still, these regular oscillations become chaotic. \cite{rlkv1,rlkv2} Here, we set $\alpha=1.95$ and initiate each neuron from fixed point initial conditions, so that the additive spatiotemporal Gaussian noise $\xi^{(i)}(n)$, having mean $<\xi^{(i)}(n)>_{i,n}=0$ and autocorrelation $<\xi^{(i)}(n) \xi^{(j)}(h)>= \delta_{ij} \delta(n-h)$, acts as the only source of large-amplitude excitations. Furthermore, in Eq.~(1) $f(x)=\frac{1}{1+x^2}$ is a nonlinear function describing the essential ingredients of neuronal dynamics, $D$ is the coupling strength, the parameter $\sigma$ determines the noise intensity, and $\tau$ is the delay length. The latter three parameters will be in the focus of attention within this work, whereas $\beta$ and $\gamma$ will be varied only occasionally.

As the base for interactions between neurons the scale-free network is employed which is generated via growth and preferential attachment as proposed by Barab\'{a}si and Albert, \cite{r21} consisting of $N=200$ vertices in this study. Each vertex corresponds to one neuron, whose dynamics is governed by the noise-driven Rulkov map. In Eq.~(1), $\varepsilon^{i,j} = 1$ if neuron $i$ is coupled to neuron $j$, and $\varepsilon^{i,j} = 0$ otherwise. The preferential attachment is introduced via the probability $\Pi$, which states \cite{r21} that a new vertex will be connected to vertex $i$ depending on its degree $k_i$ according to $\Pi(k_i )=k_i / \sum_j k_j$. This growth and preferential attachment scheme yields a network with an average degree $k_{av}=\frac{\sum_i k_i}{N}$, and a power-law degree distribution with the slope of the line equaling $\approx -3$ on a double-logarithmic graph in the present study. Notably, the analytical slope of the line is $-3$ exactly. Such networks having $k_{av}=6$ will be used throughout this work.

It remains of interest to mathematically introduce the subthreshold periodic pacemaker, which takes the form of a spike train defined by
\begin{equation}
\pi^{(r)}(n)=
\begin{cases}
g, & \mbox{if ($n$ mod $t$)} \geq (t-w),\\
0, & \mbox{else}.
\end{cases}
\end{equation}
In Eq. (2) the parameters $t$, $w$, $g$ and $n$ are defined as follows: $t$ determines the oscillation period of the spike train (pacemaker), $w$ is the width of each spike, $g$ defines the amplitude of the spikes (the baseline is $0$), and $n$ is the discrete time index introduced above in Eq. (1). Moreover, the subscript $r$ denotes the chosen neuron among all $N=200$ neurons constituting the excitable (each neuron occupies a fixed point and can thus be excited by weak perturbations) scale-free network, to which the pacemaker is introduced as an additive term to the variable $x^{(r)}(n)$. In our numerical simulations, we choose the parameter values $t=700$ (unless otherwise stated), $w=50$ and $g=0.015$, which ensure that without introducing any noise ($\sigma=0$) the pacemaker is subthreshold, meaning that by itself it cannot induce large-amplitude excitations from any of the neurons constituting the network.

To quantitatively characterize the collective response of the neuronal network, we introduce the average membrane potential $X(n)=\frac{1}{N}\sum_{i=1}^{N}x^{(i)}(n)$ as the main output to be examined further. The correlation of the average membrane potential $X(n)$ with the frequency of the pacemaker $\omega=2\pi/t$ is computed via the Fourier coefficients according to
\begin{equation}
Q_{\sin}=\frac{2}{Tt}\sum_{n=1}^{Tt}X(n)\sin(\omega n)
\end{equation}
\begin{equation}
Q_{\cos}=\frac{2}{Tt}\sum_{n=1}^{Tt}X(n)\cos(\omega n)
\end{equation}
where $T=300$ is the number of periods of the pacemaker used. Note that the sine and cosine functions in Eqs. (3) and (4) have the same frequency $\omega=2\pi/t$ as is used for pacing the neuronal network [note that $t$ in Eq. (2) is the oscillation period of the spike train], and that thus these equations determine the correlation between the output of the network $X(n)$ and the frequency of the pacemaker. We therefore use the Fourier coefficient $Q=\sqrt{Q_{\sin}^{2}+Q_{\cos}^{2}}$ as a numerically effective measure for stochastic resonance, capturing succinctly the collective spatiotemporal behavior of the neuronal network and its correlation with the pacemaker rhythm. In general, $Q$ can exhibit a bell-shaped dependence as a key parameter (for example $\sigma$) is varied, indicating the occurrence of stochastic resonance. Importantly, since the generation of scale-free networks has inherent random ingredients, which can be additionally amplified by individual vertex (neuron) pacing, final results shown below were averaged over $30$ independent runs for each set of parameter values (wherever applicable) to guarantee an appropriate accuracy.

\section{Results}

\begin{figure}
\centerline{\epsfig{file=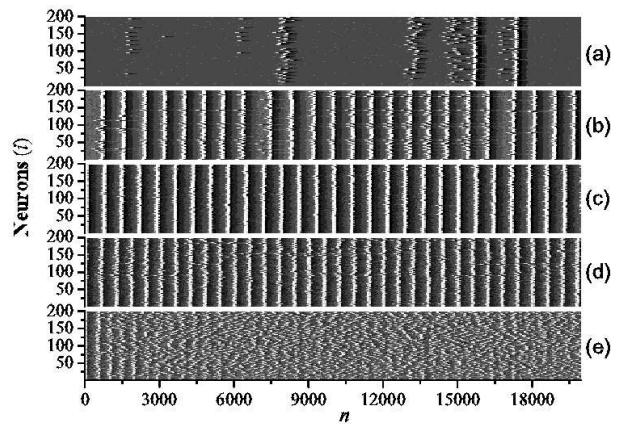,width=8cm}}
\caption{Space-time plots obtained at $\tau=0$ and $D=0.006$ for different $\sigma$ values, equaling: (a) $0.005$, (b) $0.008$, (c) $0.025$, (d) $0.05$ and (e) $0.08$. The color profile is linear, black depicting $x^{(i)}(n)=-1.7$ and white $x^{(i)}(n)=0.1$, with eight shades of gray in-between.}
\end{figure}

We start by setting $\tau=0$ and introducing the pacemaker to the neuron with the lowest degree $k_{min}$ within the network, thus $r=i(k_{min})$. Space-time plots obtained by varying $\sigma$ are presented in Fig. 1, where it can be observed that the excitatory fronts follow the pacemaker rhythm (oscillation period is $t=700$) only with an intermediate value of $\sigma$ [see Fig. 1(c)]. On the other hand, smaller $\sigma$ either largely fail to evoke excitations [see Fig. 1(a)] or the excitatory fronts have defects and are not frequent enough [see Fig. 1(b)]. Noise intensities exceeding the optimal value, however, have an ability of initiating excitations on their own (even when the pacemaker is not firing), thus again failing to conform to the weakly imposed rhythm [see Fig. 1(d)], or further still, completely overruling the deterministic dynamics [see Fig. 1(e)]. Presented results therefore indicate a classical stochastic resonance scenario, where an intermediate noise intensity ensures the best response of the system to weak external deterministic forcing. Notably, qualitatively identical results are obtained if the pacemaker is introduced to the neuron with the largest degree $r=i(k_{max})$ within the scale-free network. It is of interest to assess these observations quantitatively.

\begin{figure*}
\centerline{\epsfig{file=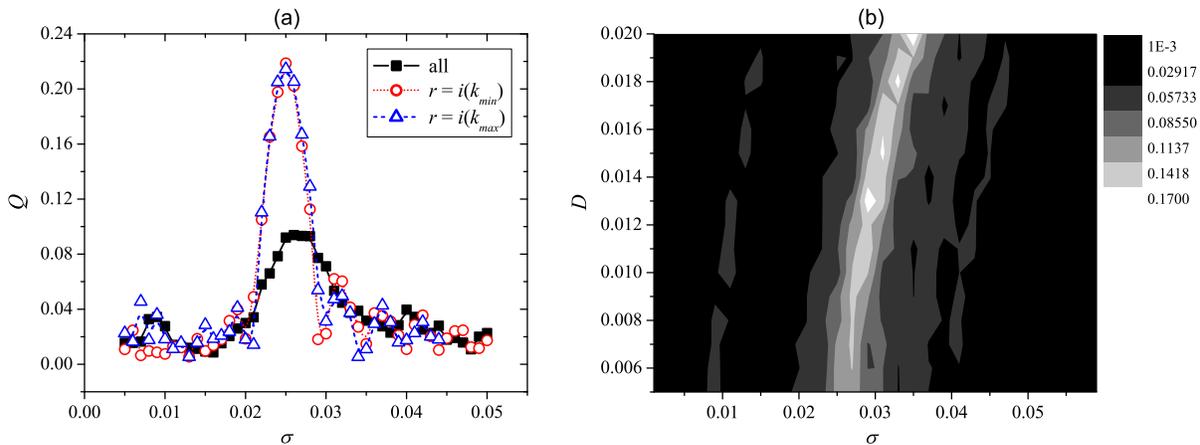,width=16cm}}
\caption{(a) Dependence of $Q$ on $\sigma$ at $\tau=0$ and $D=0.006$ for different placings of the pacemaker within the scale-free network (see also the main text for further details). (b) Dependence of $Q$ on $\sigma$ and $D$ at $\tau=0$ when the pacemaker is introduced to one of the neurons having the lowest degree [$r=i(k_{min})$].}
\end{figure*}

To describe the pacemaker-driven stochastic resonance more precisely, we consider the dependence of $Q$ on $\sigma$ in Fig. 2(a) for different options with respect to the placing of the pacemaker. It can be observed that, irrespective of whether the pacemaker is introduced to the neuron with the minimal $r=i(k_{min})$ or the maximal degree $r=i(k_{max})$, or to all neurons of the network ["all" in Fig. 2(a)], there exists an intermediate optimal noise intensity $\sigma$ for which $Q$ is maximal, thus exhibiting a bell-shaped dependence characteristic of the stochastic resonance. Interestingly, one can observe that as the pacemaker is introduced to all neurons constituting the network, the stochastic resonance decreases rather dramatically. On the other hand, whether the pacemaker is introduced to the neuron with the highest [$r=i(k_{max})$] or to one of the neurons with the lowest degree [$r=i(k_{min})$] does not notably affect the outlay and magnitude of $Q$ in dependence on $\sigma$. The placing of an individual pacemaker is mostly irrelevant because in a scale-free network chances are high that the low-degree neurons will be connected to one of the main high-degree neurons (hubs) with no more than a single direct link. Hence now, whether we introduce the pacemaker directly to the main hub or one link further away (to any of the low-degree neurons) is not vital since the pacemaker rhythm does not deteriorate much within nearest or even next-nearest neurons. Somewhat more surprising and counterintuitive is the fact that the weak periodic forcing of all neurons is less efficient in imposing a certain rhythm of excitatory fronts than an individual pacemaker. We argue that this effect is a consequence of the essentially non-biased state of neurons constituting the scale-free network, where only an individual neuron is forced. Such non-biased (not inclined towards an excitation by a pacemaker) noisy neurons can respond even to the weakest input from their neighbors, thus synchronizing optimally into practically perfect excitatory fronts. On the other hand, if all neurons are weakly forced, chances for phase slips are much higher since every neuron acts on its own (due to its individual forcing), indeed trying to enforce \textit{its} rhythm to the neighbors. Thus, competition between excitations emerges, which acts detrimental on the overall synchrony of the network, in turn decreasing the correlation with the forcing frequency and leading to the decline of stochastic resonance. In fact, the phenomenon shown in Fig. 2(a) is one of the few examples where less is actually more in terms of efficiency of noise-induced synchrony and correlation with weak external forcing in neuronal dynamics.

Before turning to the impact of finite, \textit{i.e.} non-zero delays $\tau$, we investigate in Fig. 2(b) stochastic resonance in the transition to the strong coupling region (high vales of $D$). Results are presented for $r=i(k_{min})$, but they are qualitatively identical also for $r=i(k_{max})$, and if all the neurons are exposed to the weak periodic forcing (not shown). Evidently, the stochastic resonance phenomenon prevails irrespective of $D$, except that the optimal value of $\sigma$ shifts to slightly higher values upon its increase. Thus, it can be concluded that the pacemaker-driven stochastic resonance on scale-free neuronal networks is a robust phenomenon, occurring largely independently of the particular placing of the pacemaker or the strength of the coupling. Moreover, an individually paced neuron is actually more effective in warranting ordered excitatory fronts in accordance with the weakly imposed rhythm than the global forcing of the whole network. In what follows, therefore, focus is on individually paced noise-driven scale-free neuronal networks, where one of the neurons having the lowest degree will be chosen as the input for the deterministic forcing.

\begin{figure}
\centerline{\epsfig{file=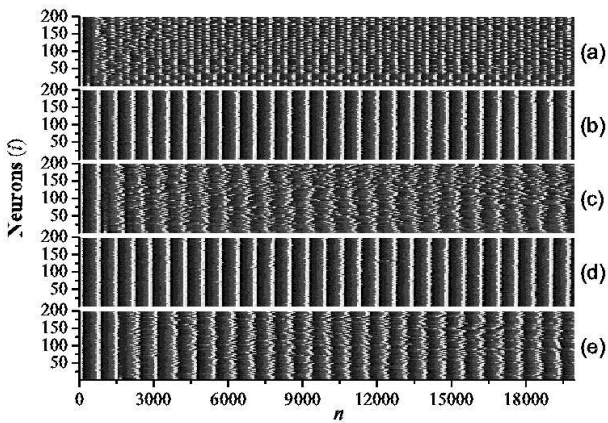,width=8cm}}
\caption{Space-time plots obtained at $D=0.006$ and $\sigma=0.025$ for different $\tau$ values, equalling: (a) $300$, (b) $700$, (c) $1000$, (d) $1400$ and (e) $1600$. In all panels the color profile is identical as in Fig. 1.}
\end{figure}

As when examining the dependence of the neuronal dynamics on $\sigma$ in Fig. 1, we start by presenting in Fig. 3 the space-time plots obtained for different values of $\tau$ while keeping the coupling strength $D=0.006$ and the noise intensity $\sigma=0.025$ fixed. Results shown in the five panels of Fig. 3 illustrate the spatiotemporal dynamics of neurons on the studied scale-free neuronal network with $r=i(k_{min})$. From careful visual inspection, an intermittent pattern of regularity and disorder can be inferred upon increasing $\tau$. In particular, while for $\tau=700$ [panel (b)] and $\tau=1400$ [panel (d)] the excitatory fronts are synchronous and largely obeying to the pacemaker rhythm, for $\tau=300$ [panel (a)], $\tau=1000$ [panel (c)] and $\tau=1600$ [panel (e)] the regularity is either completely lost or at least the excitatory fronts become ragged and gradually loosing synchrony with the imposed frequency. Indeed, the delay-induced transitions to ordered spatiotemporal dynamics on scale-free neuronal networks appear intermittently at roughly integer multiples of the period of the pacemaker, equalling $t=700$, which corresponds rather accurately to the so-called global-resonant oscillation period \cite{mmx} of each individual Rulkov neuron for the currently used parameter values $\beta = \gamma = 0.001$ and $\alpha=1.95$, which can be extracted from the map by calculating the Fourier transform of noise-driven oscillations. \cite{mmx} By setting $t$ largely different from the global-resonant oscillation period of individual neurons, however, the intermittent outlay presented in Fig. 3 can no longer be observed. Visual inspection of Fig. 3 thus reveals that regular and irregular front propagations appear intermittently as the delay is increased, indicating that finite (non-zero) delays in coupling among neurons might play a pivotal role in generating spatiotemporal order of neuronal activity on scale-free networks in accordance with a weak localized deterministic input, provided that the latter is adjusted to approximately agree with the global-resonant oscillation frequency of the neurons.

To account for the above visual interpretation quantitatively, we adjust the local dynamics of each neuron (thus far we have not varied this) by varying $\beta$ and $\gamma$, in turn affecting the speed of the temporal evolution of $y_{i}(n)$ and consequently the global-resonant oscillation frequency. At the same time, we adjust the oscillation period $t$ of the pacemaker correspondingly. In particular, we consider three different cases; namely $\beta = \gamma = 0.0006$, $\beta = \gamma = 0.001$ and $\beta = \gamma = 0.0016$, and change the oscillation period of the pacemaker to $t=1200$, $t=700$ and $t=500$, respectively. These $t$ values are in good agreement with the global-resonant frequency of an individual neuron with the corresponding values of $\beta$ and $\gamma$. Results presented in Fig. 4 show that, in accordance with the visual inspection of Fig. 3, multiple resonance in $Q$ depending upon the increase of $\tau$ are obtained by given $\sigma$ and $D$. Following the common terminology, these are termed as delay-induced multiple stochastic resonance on scale-free neuronal networks. Moreover, it is clear that the particular locations of the maxima of $Q$ shift to different values of $\tau$ as $\beta$, $\gamma$ and $t$ are varied. Crucially however, it is always so that the locking between $\tau$ and integer multiples of $t$ is preserved. Thus, resonances depending on $\tau$ appear at integer multiples of $t$ only if the latter is close to the global-resonant oscillation period \cite{mmx} of the individual neurons. On the other hand, values of $\tau$ outside the regions of multiple integers of $t$ impair the stochastic resonance significantly, as can be inferred from the rather sharp descents of $Q$ towards smaller values, as soon as the optimal $\tau$'s are replaced by other values. We therefore conclude that the delay-induced stochastic resonances of neuronal activity are due to the locking between the delay length $\tau$ and the global-resonant oscillation period of individual neurons if the latter is close to the oscillation period of the pacemaker. This is valid independently of the particular placing of the pacemaker, and also for globally paced scale-free neuronal networks with different coupling strengths.

\begin{figure}
\centerline{\epsfig{file=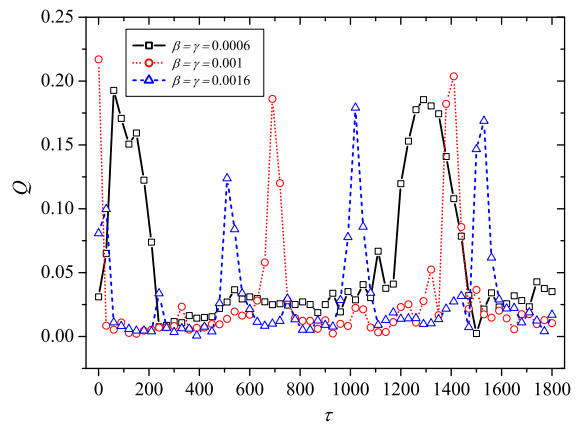,width=7.5cm}}
\caption{Dependence of $\sigma$ on $\tau$ for different combinations of $\beta$ and $\gamma$ (see also main text for further details) when the pacemaker is introduced to one of the neurons with the lowest degree [$r=i(k_{min})$]. Where applicable, other parameter values are the same as in Fig. 3.}
\end{figure}

\begin{figure*}
\centerline{\epsfig{file=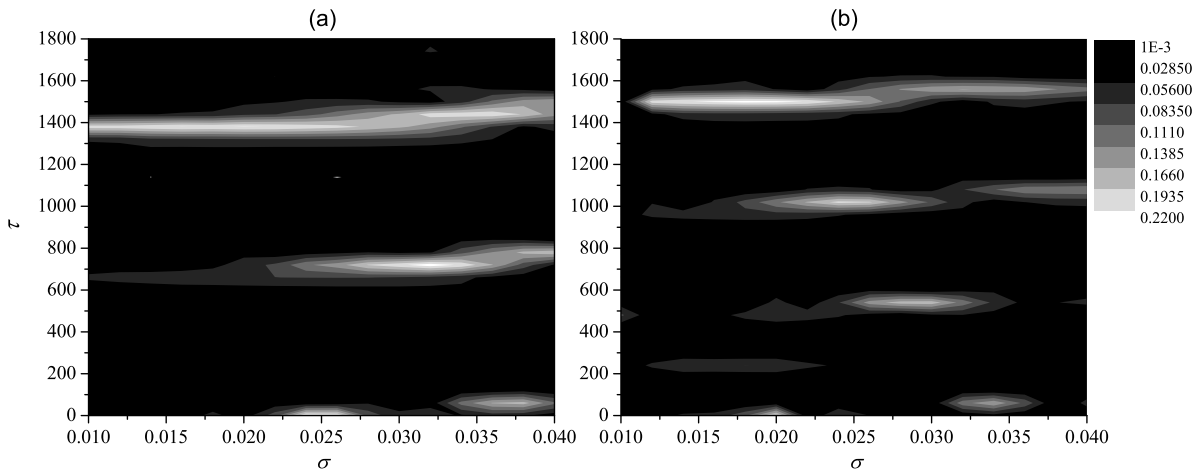,width=16cm}}
\caption{Contour plots of $Q$ in dependence on $\tau$ and $\sigma$ when the pacemaker is introduced to one of the neurons with the lowest degree [$r=i(k_{min})$] and $D=0.006$. Other parameter values are: (a) $\beta = \gamma = 0.001$, $t=700$ and (b) $\beta = \gamma = 0.0016$, $t=500$.}
\end{figure*}

Finally, we present some results where the above-outlined dependencies can be observed at a glance. Figures 5(a) and (b) show $Q$ in dependence on $\sigma$ and $\tau$ for two combinations of $\beta$ and $\gamma$. In panel (a), where $\beta = \gamma = 0.001$ and $t=700$, multiple stochastic resonances are clearly eligible as narrow white-shaded regions, appearing roughly at integer multiples of $\tau=700$ across suitable spans of $\sigma$. Practically identical, for $\beta = \gamma = 0.0016$ and $t=500$, stochastic resonances appear at integer multiples of $\tau=500$, again strengthening our argumentation with respect to the connection with the global-resonant oscillation period of individual neurons and the corresponding similarly-set forcing period of the pacemaker. It is worth emphasizing that regions of optimal $\tau$ are very narrow, especially if compared to the rather broad regions of noise intensities $\sigma$ that still warrant reasonably high $Q$, suggesting that fine-tuned delays might be crucial for efficient recognition of weak localized external signals. Also notable are some fuzzily expressed regions of stochastic resonance at $\tau \approx 70$ in both panels of Fig. 5. However, these are most likely due to the fact that shorter, if compared to the optimal, delay lengths do not influence the neuronal dynamics strong enough to fully prohibit noise-induced correlations between the pacemaker and the neuronal dynamics. This occasionally yields results similar to the case of $\tau=0$ though only at substantially higher noise intensities, which are needed to compensate for the disturbing impact of non-optimal delays. Moreover, we note that for other forcing frequencies of the pacemaker we have performed similar investigation, yet only when the pacemaker frequency is close to the global-resonant frequency \cite{mmx} of individual neurons forming the scale-free network can multiple stochastic resonances be observed at integer multiples of the forcing period.

\section{Summary and discussion}

In sum, we have studied stochastic resonance phenomena on scale-free neuronal networks in dependence on the noise intensity and time-delayed coupling when the pacemaker is introduced to different individual neurons or acts as global forcing. We found that the stochastic resonance occurs irrespective of the location of the pacemaker and of local or global forcing considerations. Yet remarkably, locally introduced weak pacemakers guarantee better expressed stochastic resonance than global forcing. Indeed, it is generally believed that global input is not common in real neuronal systems, and in fact local inputs are far more likely. In particular, given a huge number of neurons, it is unnecessary and even impossible to introduce external signals or stimuli to all. Only weak, partial and local inputs are reasonable, guaranteing low energy consumption and efficiency in large neuronal networks. Furthermore, by introducing delays to the coupling scheme, we observed multiple stochastic resonances upon fine-tuning of the delay length, which appear at every multiple of the forcing frequency if the latter is close to the global-resonant oscillation frequency of individual neurons [note that the deterministic dynamics is of fixed (equilibrium) point type]. \cite{mmx} More precisely, the multiple stochastic resonances appear in an intermittent fashion as the delay increases, where the intermittency is a direct consequence of the on/off locking between the forcing frequency and the delay length. Thus, we have shown that noise and time-delayed coupling play complementary roles in ensuring optimal detection of weak localized stimuli in scale-free neuronal networks via stochastic resonance. We therefore believe that fine-tuned delays can effectively supplement recently identified mechanisms for the enhancement of neuronal synchronization, \cite{mh1, mh2} as well as synchronization \cite{add11, add2, add3, add4} and detection of weak signals \cite{mh3} on complex and scale-free networks in general, thereby constituting an important factor of interneuronal communication. This argumentation seems to be supported also by actual biological data, demonstrating that conduction velocities along axons connecting neurons vary from 20 to 60 m/s. \cite{TS1} Real-life delays are thus within the range of milliseconds, suggesting that substantially lower or higher values may be preclusive for optimal functioning of neuronal tissues. We hope our study will be a useful supplement to the existing body of literature for the function-follow form concept, \cite{ff1, ff5} as well as the role of complex neuronal networks in general, \cite{cn1, cn3} and also serve as a viable source of information when striving towards further advances in the field.

\begin{acknowledgments}
This work was supported by the National Science Foundation of China (Grant Nos. 10702023 and 10832006) and China's Post-Doctoral Science Foundation (Grant Nos. 200801020 and 20070410022). Matja{\v z} Perc individually acknowledges support from the Slovenian Research Agency (Grant Nos. Z1-9629 and Z1-2032-2547).
\end{acknowledgments}


\begin{thebibliography}{99}

\bibitem{do1}
C. Nicolis and G. Nicolis, Tellus \textbf{33}, 225 (1981).

\bibitem{do2}
R. Benzi, A. Sutera, and A. Vulpiani, J. Phys. A \textbf{14}, L453 (1981).

\bibitem{do4}
L. Gammaitoni, P. H\"{a}nggi, P. Jung, and F. Marchesoni, Rev. Mod. Phys. \textbf{70}, 223 (1998).

\bibitem{sr1}
A. Neiman and W. Sung, Phys. Lett. A \textbf{223}, 341 (1996).

\bibitem{sr2}
S. G. Lee and S. Kim, Phys. Rev. E \textbf{60}, 826 (1999).

\bibitem{sr3}
H. E. Plesser and T. Geisel, Neurocomputing \textbf{38}, 307 (2001).

\bibitem{sr4}
K. Miyakawa, T. Tanaka, and H. Isikawa, Phys. Rev. E \textbf{67}, 066206 (2003).

\bibitem{sr5}
F. Ray and S. Sengupta, Phys. Lett. A \textbf{353}, 364 (2006).

\bibitem{cr1}
G. Hu, Phys. Rev. Lett. \textbf{71}, 807 (1993).

\bibitem{cr2}
A. Longtin, Phys. Rev. E \textbf{55}, 868 (1997).

\bibitem{cr3}
A. S. Pikovsky and J. Kurths, Phys. Rev. E  \textbf{78}, 775 (1997).

\bibitem{r20}
N. F. Rulkov, Phys. Rev. Lett. \textbf{86}, 183 (2001).

\bibitem{mtcr3}
Y. Jiang, Phys. Rev. E \textbf{71}, 057103 (2005).

\bibitem{xfx1}
C. S. Zhou, J. Kurths, and B. Hu, Phys. Rev. Lett. \textbf{87}, 098101 (2001).

\bibitem{xfx2}
F. Liu, B. Hu, and W. Wang, Phys. Rev. E \textbf{63}, 031907 (2001).

\bibitem{xfx3}
S. Wang, F. Liu, W. Wang, and Y. Yu, Phys. Rev. E \textbf{69}, 011909 (2004).

\bibitem{xfx4}
T. Kreuz, S. Luccioli, and A. Torcini, Phys. Rev. Lett. \textbf{97}, 238101 (2006).

\bibitem{xfx5}
X. Sun, Q. Lu, and J. Kurths, Physica A \textbf{387}, 6679 (2008).

\bibitem{ws}
D. J. Watts and S. H. Strogatz, Nature \textbf{393}, 440 (1998).

\bibitem{mp1}
M. Perc, Phys. Rev. E  \textbf{76}, 066203 (2007).

\bibitem{nw}
M. E. J. Newman and D. J. Watts, Phys. Lett. A \textbf{263}, 341 (1999).

\bibitem{mp2}
M. Ozer, M. Perc, and M. Uzuntarla, Phys. Lett. A \textbf{373}, 964 (2009).

\bibitem{com1}
O. Kwon and H.-T. Moon, Phys. Lett. A \textbf{298}, 319 (2002).

\bibitem{com2}
O. Kwon, H.-H. Jo, and H.-T. Moon, Phys. Rev. E \textbf{72}, 066121 (2005).

\bibitem{arr1}
C. S. Zhou, J. Kurths, and B. Hu, Phys. Rev. E \textbf{67}, 030101 (2003).

\bibitem{swsr1}
Z. Gao, B. Hu, and G. Hu, Phys. Rev. E \textbf{65}, 016209 (2001).

\bibitem{SA07}
F. Sagu\'{e}s, J. M. Sancho, and J. Garc\'{i}a-Ojalvo, Rev. Mod. Phys. \textbf{79}, 829 (2007).

\bibitem{GO99}
J. Garc\'{i}a-Ojalvo and J. M. Sancho, \textit{Noise in Spatially Extended Systems} (Springer, New York, 1999).

\bibitem{add1}
O. Carrillo, M. A. Santos, J. Garc\'{i}a-Ojalvo, and J. M. Sancho, Europhys. Lett. \textbf{65}, 452 (2004).

\bibitem{MP1}
M. Perc, Phys. Rev. E \textbf{72}, 016207 (2005).

\bibitem{or22}
Q. Y. Wang, Q. S Lu, and G. R. Chen, Eur. Phys. J. B \textbf{12}, 255 (2006).

\bibitem{r21}
A.-L. Barab\'{a}si and R. Albert, Science \textbf{286}, 509
(1999).

\bibitem{or23}
E. R. Kandel, J. H. Schwartz, and T. M. Jessell, \textit{Principles of Neural Science} (Elsevier, Amsterdam, 1991).

\bibitem{or24}
Q. Y. Wang and Q. S Lu, Chin. Phys. Lett. \textbf{3}, 543 (2005).

\bibitem{or25}
E. Rossoni, Y. H. Chen, M. Z. Ding, and J. F. Feng, Phys. Rev. E \textbf{71}, 061904 (2005).

\bibitem{or26}
Q. Wang, Z. Duan, M. Perc, and G. Chen, Europhys. Lett. \textbf{78}, 50008 (2008).

\bibitem{cn2}
V. M. Egu{\'i}luz, D. R. Chialvo, G. A. Cecchi, M. Baliki, and A. V. Apkarian, Phys. Rev. Lett. \textbf{94}, 018102 (2005).

\bibitem{sm2}
M. Kaiser, R. Martin, P. Andras, and M. P. Young, Eur. J. Neurosci. \textbf{25}, 3185 (2007).

\bibitem{rlkv1}
A. I. Shilnikov and N. F. Rulkov, Int. J. Bifurcat. Chaos \textbf{13}, 3325 (2003).

\bibitem{rlkv2}
A. I. Shilnikov and N. F. Rulkov, Phys. Lett. A \textbf{328}, 177 (2004).

\bibitem{mmx}
M. Perc and M. Marhl, Phys. Rev. E \textbf{71}, 026229 (2005).

\bibitem{mh1}
L. Huang, Y.-C. Lai, and R. Gatenby, Chaos \textbf{18}, 013101 (2008).

\bibitem{mh2}
S.-G. Guan, X.-G. Wang, Y.-C. Lai, and C. H. Lai, Phys. Rev. E \textbf{77}, 046211 (2008).

\bibitem{add11}
A. E. Motter, C. S. Zhou, and J. Kurths, Europhys. Lett. \textbf{69}, 334 (2005).

\bibitem{add2}
C. Zhou and J. Kurths, Phys. Rev. Lett. \textbf{96}, 164102 (2006).

\bibitem{add3}
A. Arenas, A. D{\'i}az-Guilera, C. J. P{\'e}rez-Vicente, Physica D \textbf{224}, 27 (2006).

\bibitem{add4}
C. Zhou, A. E. Motter, and J. Kurths, Phys. Rev. Lett. \textbf{96}, 034101 (2006).

\bibitem{mh3}
J. A. Acebr{\'o}n, S. Lozano, and A. Arenas, Phys. Rev. Lett. \textbf{99}, 128701 (2007).

\bibitem{TS1}
T. M. Sainz, C. Masoller, H. A. Braun, and M. T. Huber, Phys. Rev. E. \textbf{70}, 031904 (2004).

\bibitem{ff1}
C. C. Hilgetag, G. A. Burns, M. A. O'Neill, J. W. Scanell, and M. P. Young, Phil. Trans. R. Soc. B \textbf{355}, 91 (2000).

\bibitem{ff5}
V. Volman, I. Baruchi, and E. Ben-Jacob, Phys. Biol. \textbf{2}, 98 (2005).

\bibitem{cn1}
O. Sporns, D. R. Chialvo, M. Kaiser, and C. C. Hilgetag, Trends in Cognitive Sciences \textbf{8}, 418 (2004).

\bibitem{cn3}
C. Zhou, L. Zemanova, G. Zamora, C. C. Hilgetag, and J. Kurths, Phys. Rev. Lett. \textbf{97}, 238103 (2006).

\end{thebibliography}
\end{document}